\documentclass[12pt]{iopart}

\begin{document}
\jl{1}
\title{Comment on ``Relativistic extension of shape-invariant potentials''} 
\author{Antonio S de Castro\footnote[1]{castro@feg.unesp.br}}
\address{UNESP Campus de Guaratinguet\'a DFQ \\
Caixa Postal 205 \\
12516-410 Guaratinguet\'a SP Brasil}

\begin{abstract}
This comment directs attention to some fails of Alhaidari\'{}s approach to
solve relativistic problems. It is shown that his gauge considerations are
way off the mark and that the class of exactly solvable relativistic
problems is not so enlarged as Alhaidari thinks it is.
\end{abstract}
\submitted
\maketitle

In a recent paper to this Journal, Alhaidari \cite{alh1} presented solutions
for a number of relativistic problems. This paper is a sequel of the efforts
done in Ref. \cite{alh2} when he presented the solutions for the
Dirac-Coulomb, the Dirac-Morse and the Dirac-Oscillator problems. Alhaidari
started from a static spherically symmetric electromagnetic field and choose
a gauge such that the space component of the vector potential is given by $%
W(r)\hat{r}$ and sought for ``an alternative and proper gauge fixing
condition.'' Alhaidari\'{}s letter \cite{alh2} received a severe criticism
by Vaidya and Rodrigues \cite{vr}. In particular, the radial Dirac equation
in Vaidya and Rodrigues\'{}s comment is in contradiction with that one found
by Alhaidari. Part of the inconsistence among those equations might be
elucidated by noting that the authors use different forms for the Dirac
spinors. There is a factor $i$ multiplying either the upper or the lower
components of the spinor in order to make the radial functions real for
bound-state solutions. Alhaidari multiplied the lower component by the
factor $i$ whereas Vaidya and Rodrigues multiplied the upper component.
Nevertheless, the most serious source of contradiction arises due to an
error in the Alhaidari\'{}s radial equation. This error can be easily seen
by noting that $W(r)\hat{r}$ behaves in the same way as the momentum $%
\overrightarrow{p}$ operator under the change$\overrightarrow{r}\rightarrow -%
\overrightarrow{r}$ so that $W(r)$ should appear in the same way as $d/dr$,
namely $-W(r)+d/dr$ in the second line of the matrix equation (1). The space
component of a vector behaves in the same way as a pseudoscalar under the
space reflection transformation, however $W(r)$ is not a pseudoscalar
potential to appear behaving as the term $\kappa /r$ \cite{asc}. As a matter
of  fact, the space component of a vector can always be gauged away, in the
relativistic as well as in the nonrelativistic wave equations, and the wave
functions with and without the field just differ by  a phase factor.
Needless to say that these considerations are sufficient enough to
invalidate the gauge considerations of Alhaidari\'{}s approach.

Notwithstanding, Alhaidari\'{}s strategy for transforming the Dirac equation
into a Schr\"{o}dinger-like one is effective to solve the Dirac-Oscillator
(with a well known pseudoscalar Lorentz structure) and the
\'{}Dirac-Rosen-Morse I\'{} potential with an appropriate mixing of
pseudoscalar and vector Lorentz structures.

The spin-orbit coupling parameter $\kappa $ is defined as

\[
\kappa =\left\{ 
\begin{array}{c}
-\left( j+1/2\right) =-\left( l+1\right) ,\qquad j=l+1/2 \\ 
\\ 
+\left( j+1/2\right) =l\ \ (l\neq 0),\qquad j=l-1/2
\end{array}
\right. 
\]
so that $|\kappa |=1,2,3,...$ Thus, Alhaidari\'{}s strategy is also
effective to solve the Dirac-Coulomb problem only if $\gamma =\sqrt{\kappa
^{2}-\alpha ^{2}Z^{2}}$ can be expressed as $l\left( l+1\right) $, this is
so because the centrifugal barrier in the Schr\"{o}dinger-like equation has
the characteristic term $\kappa (\kappa +1)$. On the other hand, all the
relativistic potentials with $V=0$ presented in Ref. \cite{alh1}, such as
the Dirac-Rosen-Morse II, the Dirac-Scarf and the Dirac-P\"{o}schl-Teller
potentials, have the centrifugal barrier with the factor $\kappa (\kappa +1)$
in the corresponding Schr\"{o}dinger-like equations. Alhaidari misunderstood
the implication of restricting himself to S-wave solutions ($l=0$) thinking
that he could eliminate the centrifugal barrier. However,any integer value
of  the parameter $\kappa $ is permissible except $\kappa =0$.

To summarize, putting aside the harmful question about gauge invariance,
Alhaidari\'{}s strategy for transforming the Dirac equation into a
Schr\"{o}dinger-like does not enlarge the class of exactly solvable
potentials in the Dirac equation so much as it could appear for an
uncritical reader.

\smallskip

\ack{The author acknowledges financial support from CNPq and FAPESP.}

\smallskip

\section{Acknowledgments}

The author acknowledges financial support from CNPq and FAPESP.


\begin{thebibliography}{9}
\bibitem{alh1}  Alhaidari A D 2001 {\it J. Phys. A} \textbf{34} 9827 

\bibitem{alh2}  Alhaidari A D 2001 {\it Phys. Rev. Lett.} \textbf{87} 210405

\bibitem{vr}  Vaidya A N and Rodrigues R L 2002 hep-th/0203067

\bibitem{asc}  de Castro A S 2002 hep-th/0204004
\end{thebibliography}
\end{document}